\documentclass[twocolumn,showpacs,pra]{revtex4}
\usepackage{amssymb}
\usepackage{amsmath}
\usepackage{graphicx}
\usepackage{dcolumn}
\usepackage{bm}
\usepackage{subfigure}
\usepackage{xcolor}
\usepackage{float}

\begin{document}

\title{The best approximation of an objective state with  a given set of quantum states}
\author{Li-qiang Zhang$^1$}
\author{Nan-nan Zhou$^1$}
\author{Chang-shui Yu$^{1,2}$}
\email{ycs@dlut.edu.cn}
\affiliation{$^1$School of Physics, Dalian University of Technology, Dalian 116024, China }
\affiliation{$^2$DUT-BSU joint institute, Dalian University of Technology, Dalian 116024, China }
\begin{abstract}
Approximating a quantum state by the convex mixing of some given states has strong experimental significance and provides potential applications in  quantum resource theory. Here we find a closed form of the minimal distance in the sense of $l_2$ norm between a given $d$-dimensional objective quantum state and the state convexly mixed by those restricted in any given (mixed-) state set. In particular, we  present the minimal number of the states in the given set to achieve the optimal distance. The validity of our closed solution is further verified numerically by several randomly generated quantum states.
\end{abstract}

\maketitle

\section{Introduction}

In the past few decades, quantum information technology has been developed rapidly. The essence of quantum information processing is the preparation and the manipulation of quantum states. However, due to inherent limitations, technical or economic reasons in practical scenario, the required quantum state could not be prepared exactly as we expected. One alternative approach could be the approximate preparation of the state by convex mixing of some disposable quantum states.

In addition, the approximation of a state is also widely implied in the quantum resource theory. As we know, the quantification of quantum features is the core of the resource theory. Many important quantum features such as quantum entanglement \cite{B1,B2,B3,o1,o2,o3}, quantum coherence \cite{c1,c2,c3,c4,c5,c6}, quantum discord \cite{d1,d2,d3,d4,d5,d6} and so on \cite{t1,t2} has been quantitatively studied from the point of resource theory of view. One of the most common methods  is to measure the nearest distance between the target state and the free state set \cite{n1,n2,n8}. For example, the entanglement measure can be quantified by the smallest distance between the target state and the separable state set \cite{en1,en2,en3}. Quantum discord of a quantum state can be measured by its closest distance from the set of classically correlated quantum states \cite{dis01,dis1,dis2,dis3,dis4,dis5,dis6,dis7,dis8,dis9}. Quantum coherence can be described based on the minimal distance between the target quantum state and the convex combinations of the given orthogonal basis \cite{co03,co1,co3,co4,co5,co6,co7}. Quantum superposition measures the nearest distance between the given state and some linearly independent states\cite{s1,s2,s3}. Therefore, a most general question  extracted is how to optimally approximate an objective quantum state by  the convex mixing of some given quantum states.

The  optimal approximation of a quantum state with limited states has been addressed in various cases \cite{CC1,CC2,CC3,CC4,CC5,CC6}. In Refs. \cite{CC1,CC2}, optimally approximating an unavailable quantum state $\rho$ (quantun channel $\Phi$) by the convex mixing of states (channels) drawn from a set of available states $\left \{v_{i} \right \}$ (channels $\left \{\Psi_{i} \right \}$) was considered. The approximation of a state by the  six eigenstates of Pauli matrices  was studied in Ref. \cite{CC2}  based on the $B_3$-distance, which was further revised and supplemented in Ref. \cite{CC3}. The $B_2$-distance  and the trade-off relations of sum and squared sum were investigated in Ref. \cite{CC4}. Later, the disposable quantum state set was extended from the eigenstates of the Pauli matrix to the eigenstates of any quantum logic gate \cite{CC5}, and then to arbitrary quantum states without any restriction \cite{CC6}. Up to now all the relevant contributions have been only restricted to the qubit states.

In this paper, we study the optimal approximation of a general $d$-dimensional quantum state by convex mixing the states in a given state set.  We employ the  $l_{2}$ norm to measure the distance between two quantum states. With any given state set, we give the closed solution to the question, that is, we find the minimal distance  between the objective state and the optimal state mixed with the states in the set. In particular, we can give the minimal number of the states in the set to achieve the optimal distance. We also prove that the case with the state set including more than $d^2$  states can be transformed into the case with the set including no more than $d^2$  states. In order to validate our closed solution, we investigate several examples with different dimensions in the numerical way.  All the examples demonstrates the perfect consistency with our closed solution.
The remaining of this paper is organized as follows. In Sec. II, we give a brief description of the question of the convex approximation question and the closed solution to the question. In Sec. III, we consider several randomly generated examples to test our closed results. The  discussion and conclusion  are given in Sec. IV.

\section{The approximation of the given objective state}

Let $\rho $ denote an objective state and $S:=\left\{\rho_{i} ,i=1,2,\cdots ,N\right\} $ denote a given set of quantum states. Our goal is to prepare a quantum state $%
\sigma =\chi _{1,2,\cdots ,K}\left( \vec{p}\right)
=\sum\limits_{i=1}^{K}p_{i}\rho_{i}$ with the subscripts of $\chi$ in increasing order  by the convex mixing of $K\leq N$ quantum states in
the set $S$ so that the distance between the objective quantum state $\rho $ and prepared quantum state $\sigma $ is the closest. For convenience, we'd like to consider all the states in
the representation (labelled by $'X'$) defined by some Hermitian matrix basis, e.g. $\left\lbrace X_{i}:i=0,1,2,\cdots, d^{2}-1\right\rbrace $  with $X_{0} =\mathbb{I}_{d}/\sqrt{d} $, $X_i^\dag=X_i$,  $ Tr X_{i}^\dag X_{j} =\delta_{ij} $. Thus any a $d$-dimensional quantum state $\rho$ in the $X$ representation can be expanded as $\rho=\sum_{i=0}^{d^{2}-1} r_{i}X_{i}$. The Bloch representation is such a typical example in which case $r_{i}$ is the element of the Bloch vector.
Similarly, we use $\mathbf{r}_{o}$ to represent the vector of $\rho $ in the $X$ representation with its elements $%
r_{oi }=\mathrm{Tr}(\rho X_{i})$ and $\mathbf{r}_{k}$ ($r_{ki }$ is its element) to denote the vector of the $k$th state in $S$.
In this sense, the distance between our objective state $\rho$ and the state $\sigma$ to be prepared can be given based on  the $l_{2}$ norm as
\begin{equation}
D\left( \rho ,\sigma \right) =\frac{1}{2}\left\Vert \mathbf{r_{o}} -\sum_{i=1}^{K}p_{i}\mathbf{r_{i}} \right\Vert _{2}^{2},\label{D}
\end{equation}%
with $\left\Vert\mathbf{r}\right\Vert _{2}=\sqrt{\mathbf{r}^{T}\mathbf{r}}$.
It is obvious that $D\left( \rho
,\sigma \right) =0$ for $\rho =\sigma $ and $D\left( \rho ,\sigma \right) =1$
for $\rho \perp \sigma $.

To construct the optimal quantum state $\sigma$ which is the closest to the state $\rho$ is equivalent to achieve $\min_{\vec{p}}D\left( \rho
,\chi _{1,2,\cdots ,N}\left( \vec{p}\right) \right) $. This minimization is a global convex optimization problem  for unconstrained $\vec{p}$ \cite{T1,T2,T3,T4}, which can be verified by
the Hessian matrix of this problem  defined by \begin{equation}
H=\frac{\partial ^{2}D}{\partial \vec{p}^{2}}=\mathcal{ R}_K^{T}\mathcal{R}_K,  \label{hessian}
\end{equation}%
with $\mathcal{R}_K=\left( \mathbf{r}_{1},\mathbf{r}_{2},\cdots ,\mathbf{r}%
_{K}\right) $ being a $d^{2}\times K$ matrix ($K\leq N$), and the linear constraint $\sum_{i=1}^{K}p_{i}-1=0$ and the inequality convex constraint $-p_{i}\leq0$.  The global convex optimization provides a good property: The optimal solution can be obtained by the local optimal point if it satisfies the constraints, or at the constraint boundary if the local optimal point is not within the constraint range. With these knowledge, we can present our main results about the best approximation of the objective state as follows.

\textbf{Theorem 1}.- Given an objective state $\rho$ and a state set $S$ composed of $N$ quantum states $\rho_i$ ($N\leq d^{2}$), in the $X$ representation, one can define  the vector $\mathbf{B}$ as $\mathbf{B}(i)=\left(\mathbf{r}_{i}-\mathbf{r}_{K}\right)^{T}\mathbf{r}_{o}+\delta_{iK}$  and  a matrix $A$ as $A(i,j)=\left(\mathbf{r}_{i}-\mathbf{r}_{K}\right)^{T}\mathbf{r}_{j}+\delta_{iK}$ with $K\leq N$  and $i,j=1,2,3,...,N$. The minimal distance between $\rho$ and $\sigma=\sum_{i=1}^{K}p_i\rho_i$ is given by
\begin{equation}
\min_{K\leq N}D(\rho,\chi_{i_{1},i_{2},...,i_{K}}\left( \vec{p}\right)),
\end{equation}
 where $ \vec{p}=A^{-1}\mathbf{B}$
with $\mathrm{rank}(A)=K$ and $p_i>0$ required, and  $i_{\alpha}$ denotes the $\alpha$th element in the subset composed of $K$ states from the set $S$.

\textbf{Proof}. For $N\leq d^{2}$, Eq. (\ref{D}) can be rewritten as%
\begin{equation}
D(\rho ,\sigma )=\frac{1}{2} \sum_{ij}^{N}\left( p_{i}p_{j}%
\mathbf{r}_{i}^{T}\mathbf{r}_{j}-2p_{i}\mathbf{r}_{o}^{T}\mathbf{r}_{i}+%
\mathbf{r}_{o}^{T}\mathbf{r}_{o}\right) .  \label{d1}
\end{equation}%
Consider the Lagrangian function%
\begin{eqnarray}
L(p_{i},\lambda ,\lambda _{i})=D(\rho ,\sigma )-\sum_{i}^{N}\lambda _{i}p_{i}+\lambda \left( \sum_{i}^{N}p_{i}-1\right) ,  \label{l1}
\end{eqnarray}%
where $\lambda $ and $\lambda _{i}$ are the Lagrangian multipliers. The
Karush-Kuhn-Tucker conditions \cite{KKT} are given by%
\begin{eqnarray}
\frac{\partial L}{\partial p_{i}} &=& \sum_{j}^{N}\mathbf{r}_{j}^{T}\mathbf{r}_{i}p_{j}-\mathbf{r}_{o}^{T}\mathbf{r}_{i}-\lambda _{i}+\lambda =0,  \notag \\
\lambda _{i}p_{i} &=&0,\lambda _{i}\geq 0,p_{i}\geq 0,\sum_{j}^{N} p_{j}-1=0,\notag \\
i&=&1,2,3,...,N.\label{l2}
\end{eqnarray}

After eliminating $\lambda$ by $\frac{\partial L}{\partial p_{i}}-\frac{\partial L%
}{\partial p_{N}}=0$, we have%
\begin{eqnarray}
\sum_{j}^{N}\left( p_{j}\mathbf{r}_{j} \right)^{T}\left(\mathbf{r}_{i}-\mathbf{r}_{N}\right)&=&\mathbf{r}_{o}^{T}\left(\mathbf{r}_{i}-\mathbf{r}_{N}\right)+\lambda_{i}-\lambda_{N}  \notag\\
i&=&1,2,3,...,N-1.  \label{l3}
\end{eqnarray}%
For convenience, we first consider the case all $p_{i}\neq0$ which mean $\lambda_{i}=0$ for $i=1,2,3,...,N$.
We rewrite Eq. (\ref{l3}) above in matrix form $A\mathbf{P}=\mathbf{B}$ with $\mathbf{P}(i)=\tilde{p}_{i}$ for $i=1,2,3,...,N$. The determinant of matrix $A$ is
\begin{equation}
\mathrm{det}(A)=\mathrm{det}(\mathcal{R}_{-}^{T}\mathcal{R}_{-})
\end{equation}
with $\mathcal{R}_{-}=\left( \mathbf{r}_{1}-\mathbf{r}_{N},\mathbf{r}_{2}-\mathbf{r}_{N},\cdots ,\mathbf{r}%
_{N-1}-\mathbf{r}_{N}\right) $ being a $d^{2}\times (N-1)$ matrix.

\textit{Case 1}: $\mathrm{det}(A)=0$. This means that each column of the matrix $\mathcal{R}_{-}$ is linearly related. Any quantum state represented by $\mathbf{r}_{o}=\sum_{i=1}^{N}p_{i}\mathbf{r}_{i}$ with $p_{i}\in \lbrack 0,1]$ and $\sum_{i}^{N}p_{i}=1$ must be represented by $\mathbf{r}_{o}=\sum_{j=1}^{N-1}q_{j}\mathbf{r}_{j}$ with $q_{j}\in \lbrack 0,1]$ and $\sum_{j}^{N-1}q_{j}=1$, which is implied by the Caratheodory theorem or explicitly shown in our latter theorem 2. In other words, the optimal solution of $N$ quantum states is equivalent to that of $N-1$ quantum states. The optimal distance is given by
\begin{equation}
\min_{i_{1}<i_{2}<...<i_{N-1}}D(\rho,\chi_{i_{1},i_{2},...,i_{N-1}\left( \vec{p}\right)}),i_{\alpha}=1,2,3,...,N-1.
\end{equation}

\textit{Case 2}: $\mathrm{det}(A)\neq0$. One can first calculate \begin{equation}
\mathbf{P}=A^{-1}\mathbf{B}. \label{l5}
\end{equation}
If all $\tilde{p}_{i}\in \lbrack 0,1]$ in $\mathbf{P}$, then the optimal weights $p_{i}=\tilde{p}_{i}$. By substituting the optimal weights $p_{i}$ into the prepared quantum state $\chi _{1,2,\cdots ,N}\left( \vec{p}\right)=\sum_{i}^{N}p_{i}\rho_{i}$, we can obtain the optimal distance $D(\rho,\chi _{1,2,\cdots ,N}\left( \vec{p}\right))$. If not all $\tilde{p}_{i}\in \lbrack 0,1]$, it means that the optimal weights should be at the boundary, that is, at least one of the probabilities $p_{i}$ is 0. Thus one need to consider the mixing of $N-1$ states and the optimization problem is converted to
\begin{equation}
\min_{i_{1}<i_{2}<...<i_{N-1}}D(\rho,\chi_{i_{1},i_{2},...,i_{N-1}\left( \vec{p}\right)}),i_{\alpha}=1,2,3,...,N-1.\label{l66}
\end{equation}
Repeating the process from Eq. (\ref{l3}) to Eq. (\ref{l66}), until  the probabilities $\tilde{p}_{i}\in \lbrack 0,1]$. Suppose the first valid probability vector $\mathbf{P}$ is found when considering the mixing of $M$ states, the optimal distance is taken as the minimal distance over all $C^M_N$ combinations of the $M$ states with $M\leq N$. The proof is completed.
    \hfill $\blacksquare $

\textbf{Theorem 2}.- If there are $N> d^{2}$ states in the set $S$, the optimization approximation is determined by
\begin{equation}
\min_{i_{1}<i_{2}<\cdots<i_{d^{2}}}D(\rho,\chi_{i_{1},i_{2},...,i_{d^{2}}}\left( \vec{p}\right)),i_{\alpha}=1,2,3,\cdots,d^{2}. \label{theo3}
\end{equation}

\textbf{Proof}. In the $X$ representation, the prepared states $\sigma$ can be expressed as
$
\mathbf{r}_{\sigma}=\sum_{i=1}^{N}p_{i} \mathbf{r}_{i}.
$
 Caratheodory theorem \cite{Ca1,Ca2} shows that $\mathbf{r}_{\sigma}$ can be represented by the convex combination of no more than $d^{2}+1$ vectors in the set $\tilde{S}:= \left\lbrace \mathbf{r}_{i}|i=1,2,3,...,N \right\rbrace $ such as
$
\mathbf{r}_{\sigma}=\sum_{j=1}^{d^{2}+1}q_{j} \mathbf{r}_{j},
$
with $q_{j}\geq 0$ and $\sum_{j=1}^{d^{2}+1}q_{j}=1$.

Considering that $d^{2}+1$ vectors in $\tilde{S}$ must be linearly independent, there exist $l_{i}, i=1,2,\cdots, d^{2}+1$ such
 that $\sum_{i=1}^{d^{2}+1} l_{i}\mathbf{r}_{i} =0$, which implies  $\sum_{i=1}^{d^{2}+1}
l_{i}=0$ due to $X_{0} =\mathbb{I}_{d}/\sqrt{d} $. Thus one can obtain
\begin{equation}
\mathbf{r}_{\sigma}=\sum_{i=1}^{d^{2}+1}q_{i}\mathbf{r}_{i} -\alpha \sum_{i=1}^{d^{2}+1} l_{i}\mathbf{r}_{i}
=\sum_{i=1}^{d^{2}+1}q_{i}(1-\alpha\frac{l_i}{q_i})\mathbf{r}_{i}. \label{top}
\end{equation}%
Let $\alpha=\frac{q_{i'}}{\tilde{l}_{i'}}=\min_{1\leq i \leq d^{2}+1}\left\lbrace \frac{q_i}{\tilde{l}_i}|l_{i}> 0 \right\rbrace$, we will find that $1-\alpha\frac{l_{i}}{q_{i}}\left \{
\begin{array}{cc}
  =0&   i=i'  \\
 \geq 0 &i\neq i'
    \end{array}
\right.$ and $\sum_i q_i(1-\alpha\frac{l_{i}}{q_{i}})=1$ which mean that at most $d^{2}$ vectors in the set $\tilde{S}$ are enough to convexly construct $\mathbf{r}_\sigma$.

It implies that for $N> d^{2}$, the convex mixing of only $d^{2}$ states
in $S$ is enough to achieve the optimal distance. Therefore, we can directly
consider all potential combinations of only $d^{2}$ quantum states among the set $S$%
. The minimal distance will give our expected optimal result. The proof is finished. 				\hfill $\blacksquare $

\section{Examples}

To verify the reliability of our theorems, we provide several randomly generated density matrices and compare our closed analytic results with the numerical results. In the following, the objective state $\mathbf{r}_{o}$ in the $X$ representation is given as
\begin{equation}
\mathbf{r}_{o}=k\mathbf{r}_{o1}+(1-k)\mathbf{r}_{o2},k\in[0,1],
\end{equation}
where  $\mathbf{r}_{01}$ is given in several special cases and  $\mathbf{r}_{02}$ will be randomly generated by Matlab.
 The explicit expression of $N$ quantum states in set $S=\left\{ \mathbf{r}_{1},\mathbf{r}_2 ,\cdots ,\mathbf{r}_N\right\} $ in all the below examples are given in  Appendix A.

(i)$d=2$ and $N=3$. According to theorem 1, we first consider two states, $\mathbf{r}_{1}$  and $\mathbf{r}_{2}$. The pseudo probability reads
\begin{eqnarray}
\tilde{p}_{1}&=&\frac{\left( \mathbf{r}_{o}-\mathbf{r}_{2} \right)^{T}\left( \mathbf{r}_{1}-\mathbf{r}_{2} \right)}{\Vert\mathbf{r}_{1}-\mathbf{r}_{2}\Vert_{2}^{2}} ,  \notag \\
\tilde{p}_{2}&=&1-\tilde{p}_{1}.\label{P2}
\end{eqnarray}%
With all the three states in the set taken into account, the pseudo probability reads
\begin{eqnarray}
\tilde{p}_{1} &=&\frac{1}{d}(\mathbf{r}_{1}-\mathbf{r}_{2})^{T}\left[  (\mathbf{r}_{o}-\mathbf{r}_{2})(\mathbf{r}_{2}-\mathbf{r}_{3})^{T} \right. \notag \\
&&-\left.(\mathbf{r}_{2}-\mathbf{r}_{3})(\mathbf{r}_{o}-\mathbf{r}_{2})^{T}\right] (\mathbf{r}_{2}-\mathbf{r}_{3}) , \notag\\
\tilde{p}_{2} &=&\frac{1}{d}(\mathbf{r}_{1}-\mathbf{r}_{2})^{T}\left[  (\mathbf{r}_{1}-\mathbf{r}_{3})(\mathbf{r}_{o}-\mathbf{r}_{1})^{T} \right. \notag \\
&&-\left.(\mathbf{r}_{o}-\mathbf{r}_{1})(\mathbf{r}_{1}-\mathbf{r}_{3})^{T}\right] (\mathbf{r}_{1}-\mathbf{r}_{3}) ,  \notag \\
\tilde{p}_{3} &=&1-\tilde{p}_{1}-\tilde{p}_{2},  \label{P3}
\end{eqnarray}%
where
\begin{eqnarray}
d =\Vert\mathbf{r}_{1}-\mathbf{r}_{2}\Vert_{2}^{2}\Vert\mathbf{r}_{3}-\mathbf{r}_{2}\Vert_{2}^{2}-\left[(\mathbf{r}_{1}-\mathbf{r}_{2})^{T}(\mathbf{r}_{3}-\mathbf{r}_{2}) \right] ^{2}.  \label{dd}
\end{eqnarray}
Collecting the cases with all $p_i=\tilde{p}_{i} \geq 0$, one can find the minimal distance in terms of
\begin{eqnarray}
D\left( \rho ,\sigma \right) =\frac{1}{2}\left\Vert \mathbf{r_{o}} -\sum_{i=1}^{N}p_{i}\mathbf{r_{i}} \right\Vert _{2}^{2}.
\end{eqnarray}
For example, we can make
\begin{eqnarray}
\mathbf{r}_{01}^{1}&&=(\begin{array}{cccc}
\dfrac{1}{\sqrt{2}} ,& 0, & 0, & 0  %
\end{array}%
\mathbf{)^{\intercal }},
\end{eqnarray}
which is the maximally mixed state. Then three different kinds of $\mathbf{r}_{02}$ are randomly generated as
\begin{eqnarray}
\mathbf{r}_{02}^{1}&&=(\begin{array}{cccc}
1/\sqrt{2} ,& -0.0989 &   0.1337 &  -0.1564  %
\end{array}%
\mathbf{)^{\intercal }}, \notag\\
\mathbf{r}_{02}^{2}&&=(\begin{array}{cccc}
1/\sqrt{2} ,& -0.1810  &  0.0522  &  0.2173  %
\end{array}%
\mathbf{)^{\intercal }}, \notag\\
\mathbf{r}_{02}^{3}&&=(\begin{array}{cccc}
1/\sqrt{2} ,&   0.2285 &  -0.0403   & 0.2218  %
\end{array}%
\mathbf{)^{\intercal }}.
\end{eqnarray}
The optimal distance denoted by $D(\rho)$ versus $k\in \lbrack 0,1]$ is plotted in Fig. 1 (a), where the dotted blue, solid red and dashed green lines correspond to $\mathbf{r}_{02}^{1}$, $\mathbf{r}_{02}^{2}$ and $\mathbf{r}_{02}^{3}$, respectively. The figure validates our theorem 1 based on the perfect consistency. In Fig. 1 (b), one can find that for qubit states $\mathbf{r}_{02}^{2}$ and $\mathbf{r}_{02}^{3}$, only one quantum state is enough to achieve the optimal distance when $k>0.33$.

\begin{figure}[tbp]
\centering
\subfigure[]{\includegraphics[width=0.50\columnwidth,height=1.3in]{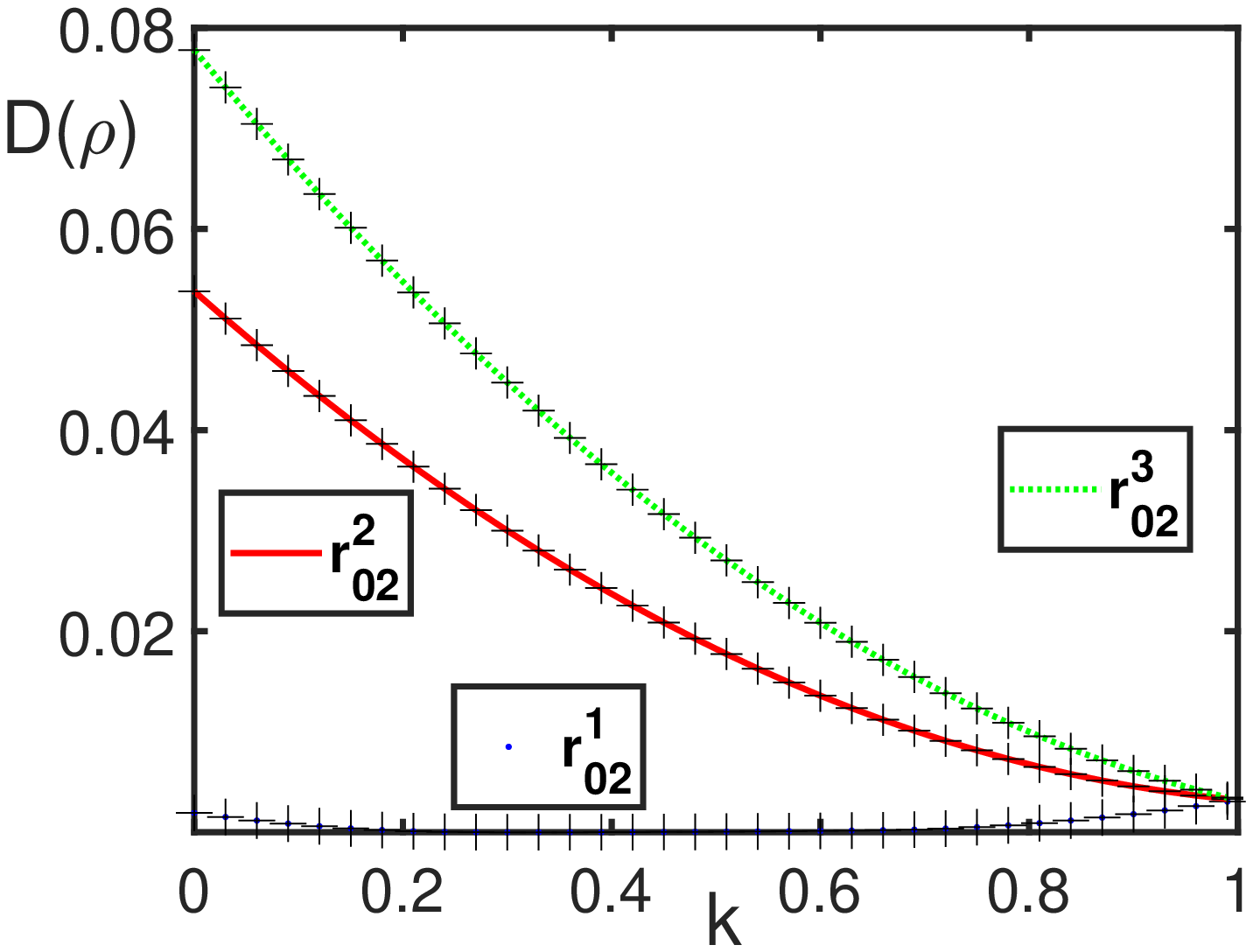}}
\subfigure[]{\includegraphics[width=0.46\columnwidth,height=1.3in]{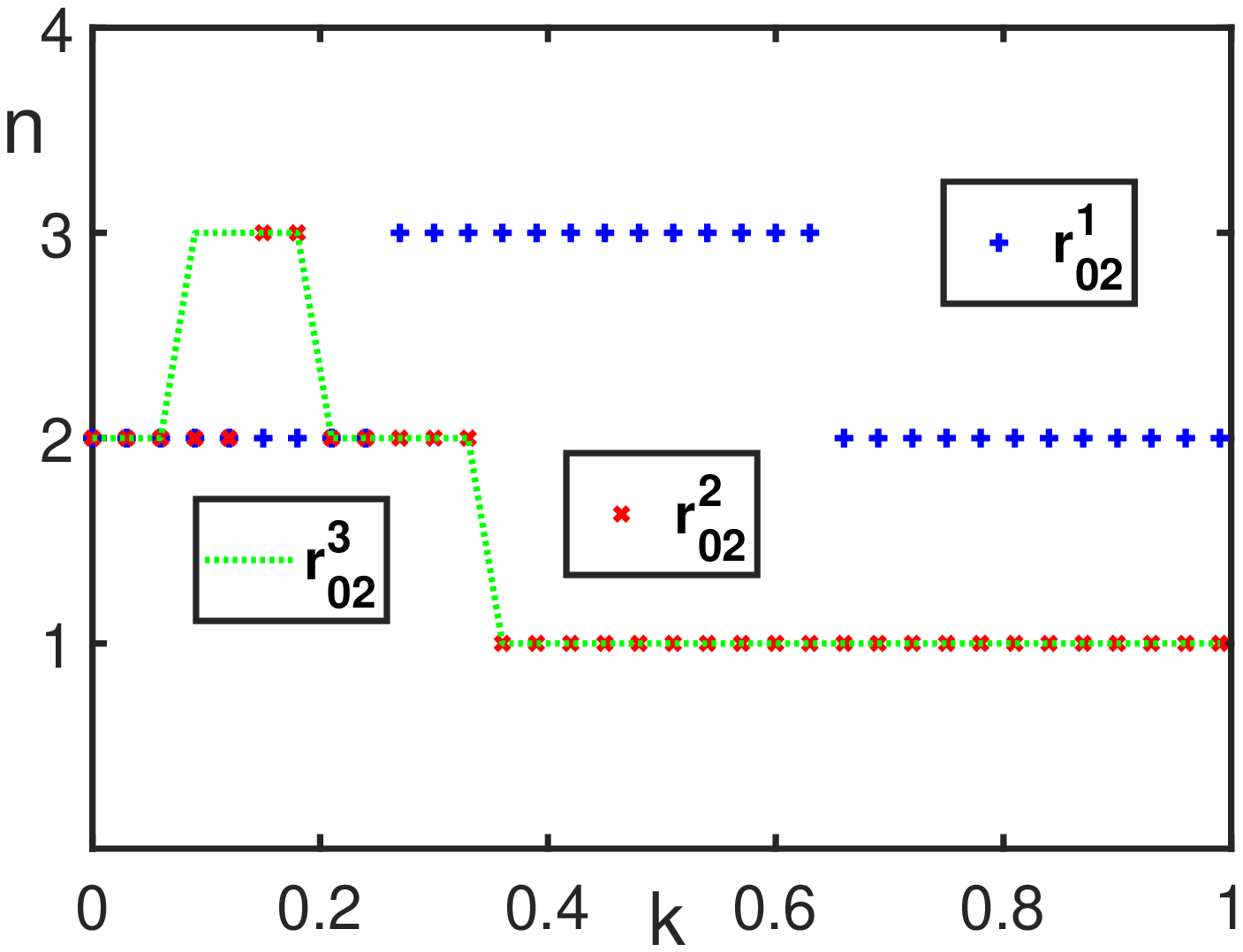}}
\caption{(color online) The optimal distance $D(\protect\rho )$ versus
various parameters $k$ in (a) for $d=2$ and $N=3$. The solid line corresponds to the strictly
closed expressions given in Eq. (\ref{P2}) and Eq. (\ref{P3}), while the numerical solutions
are marked with "+". The minimum number $n$ of quantum states in set $S$ needed for the optimal approximation of each objective state is shown in (b).}
\end{figure}

(ii)$d=2$ and $N=6$. We let $\mathbf{r}_{o1} $ take three choices:
\begin{eqnarray}
\mathbf{r}_{01}^{1}&&=(\begin{array}{cccc}
\dfrac{1}{\sqrt{2}} ,& 0, & 0, & 0  %
\end{array}%
\mathbf{)^{\intercal }}, \notag\\
\mathbf{r}_{01}^{2}&&=(\begin{array}{cccc}
\dfrac{1}{\sqrt{2}} ,& \dfrac{1}{\sqrt{6}} ,& \dfrac{1}{\sqrt{6}} ,& \dfrac{1}{\sqrt{6}}  %
\end{array}%
\mathbf{)^{\intercal }}, \notag\\
\mathbf{r}_{01}^{3}&&=(\begin{array}{cccc}
\dfrac{1}{\sqrt{2}} ,& \dfrac{1}{\sqrt{4}} ,& -\dfrac{1}{\sqrt{4}} ,& 0  %
\end{array}%
\mathbf{)^{\intercal }}.
\end{eqnarray}
Here $\mathbf{r}_{02}$ is randomly generated as
\begin{equation}
\mathbf{r}_{02}=(\begin{array}{cccc}
1/\sqrt{2} ,& 0.4533 ,& 0.1255 ,& 0.5061  %
\end{array}%
\mathbf{)^{\intercal }}.
\end{equation}
The optimal distance denoted
by $D(\rho)$ versus $k\in \lbrack 0,1]$ is plotted in Fig. 2 (a), where the dotted blue, solid red and dashed green lines correspond to $\mathbf{r}_{01}^{1}$, $\mathbf{r}_{01}^{2}$ and $\mathbf{r}_{01}^{3}$, respectively. It is shown that our closed analytic solution is completely consistent with the numerical solution. In Fig. 2 (b), we can find that for the optimal approximation of the objective quantum state, the minimal number of quantum states in set $S$ is up to 4.

\begin{figure}[tbp]
\centering
\subfigure[]{\includegraphics[width=0.50\columnwidth,height=1.3in]{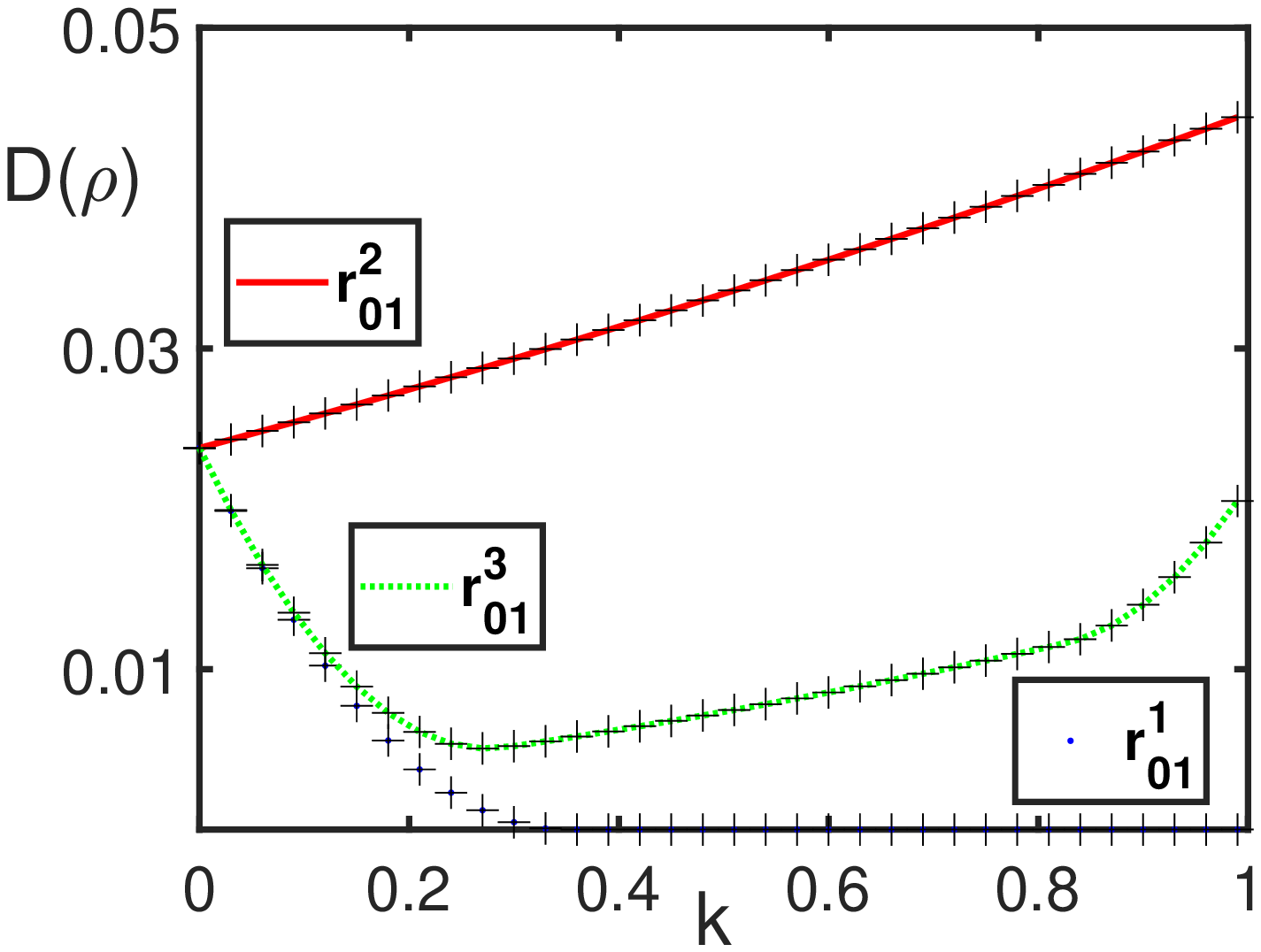}}
\subfigure[]{\includegraphics[width=0.46\columnwidth,height=1.3in]{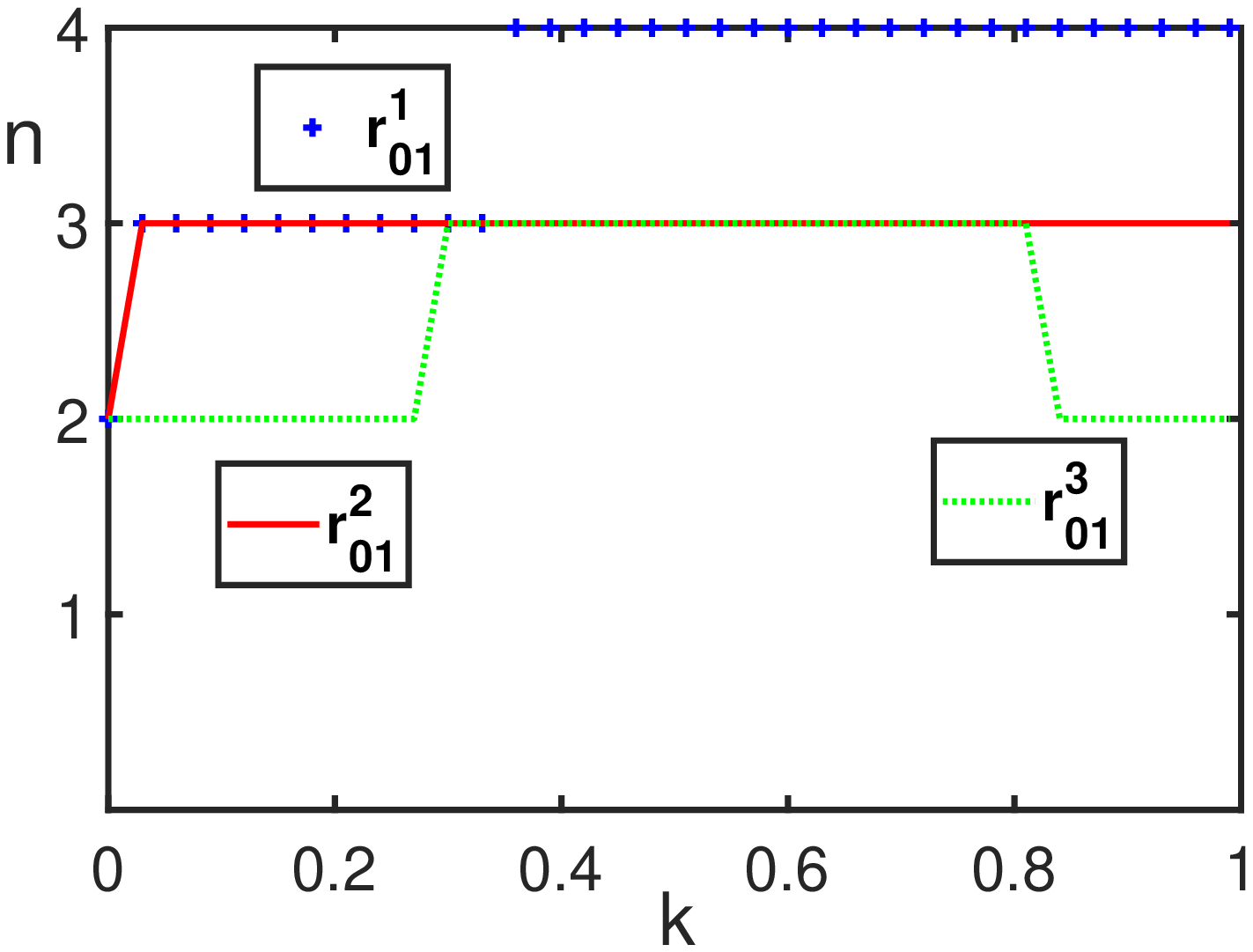}}
\caption{(color online) The optimal distance $D(\protect\rho )$ versus
various parameters $k$ in (a) for $d=2$ and $N=6$. The solid line corresponds to the strictly
closed expressions given in our theorems, while the numerical solutions
are marked with "+". The minimum number $n$ of quantum states in set $S$ needed for the optimal approximation of each objective state is shown in (b).}
\end{figure}

(iii)$d=3$ and $N=15$. In this case, $\mathbf{r}_{02}$ is randomly generated as
\begin{eqnarray}
\mathbf{r}_{02}= && \left( \begin{array}{ccccc}
1/\sqrt{3} ,& 0.0568 ,& 0.1463 ,& 0.1405  ,& -0.0456,
\end{array} \right.  \notag \\
&& \left. \begin{array}{cccc}
  -0.0531 ,& -0.1342 ,& 0.1669 ,& -0.0918  %
\end{array}%
\right)^{\mathbf{\intercal}}.
\end{eqnarray}
and
the three special quantum states are considered for $\mathbf{r}_{01}$ as
\begin{eqnarray}
\mathbf{r}_{01}^{1}=&&(\begin{array}{ccccccccc}
\dfrac{1}{\sqrt{3}} ,& 0 ,& 0 ,& 0 ,& 0 ,& 0 ,& 0 ,& 0 ,& 0 %
\end{array}%
\mathbf{)^{\intercal }}, \notag\\
\mathbf{r}_{01}^{2}=&&(\begin{array}{ccccccccc}
1 ,& \dfrac{1}{2}  ,& \dfrac{1}{2} ,& \dfrac{1}{2} ,& \dfrac{1}{2} ,& \dfrac{1}{2} ,& \dfrac{1}{2} ,& \dfrac{1}{2} ,& \dfrac{1}{2}  %
\end{array}%
\mathbf{)^{\intercal }}/\sqrt{3}, \notag\\
\mathbf{r}_{01}^{3}= && \left( \begin{array}{ccccc}
\sqrt{\frac{1}{3}} ,& \sqrt{\frac{1}{6}} ,& \sqrt{\frac{2}{6}} ,& \sqrt{\frac{3}{6}}  ,& \sqrt{\frac{4}{6}},
\end{array} \right.  \notag \\
&& \left. \begin{array}{cccc}
\sqrt{-\frac{5}{6}} ,& \sqrt{-\frac{6}{6}} ,& \sqrt{-\frac{7}{6}},& \sqrt{-\frac{8}{6}}  %
\end{array}%
\right)^{\mathbf{\intercal}}.
\end{eqnarray}
 The optimal distance denoted
by $D(\rho)$ versus $k\in \lbrack 0,1]$ is plotted in Fig. 3 (a), which validates our theorem based on the perfect consistency. The Fig. 3 (b) shows that the minimal number of quantum states in set $S$ used to optimally approximate the objective quantum state is no more than 9.

\begin{figure}[tbp]
\centering
\subfigure[]{\includegraphics[width=0.50\columnwidth,height=1.3in]{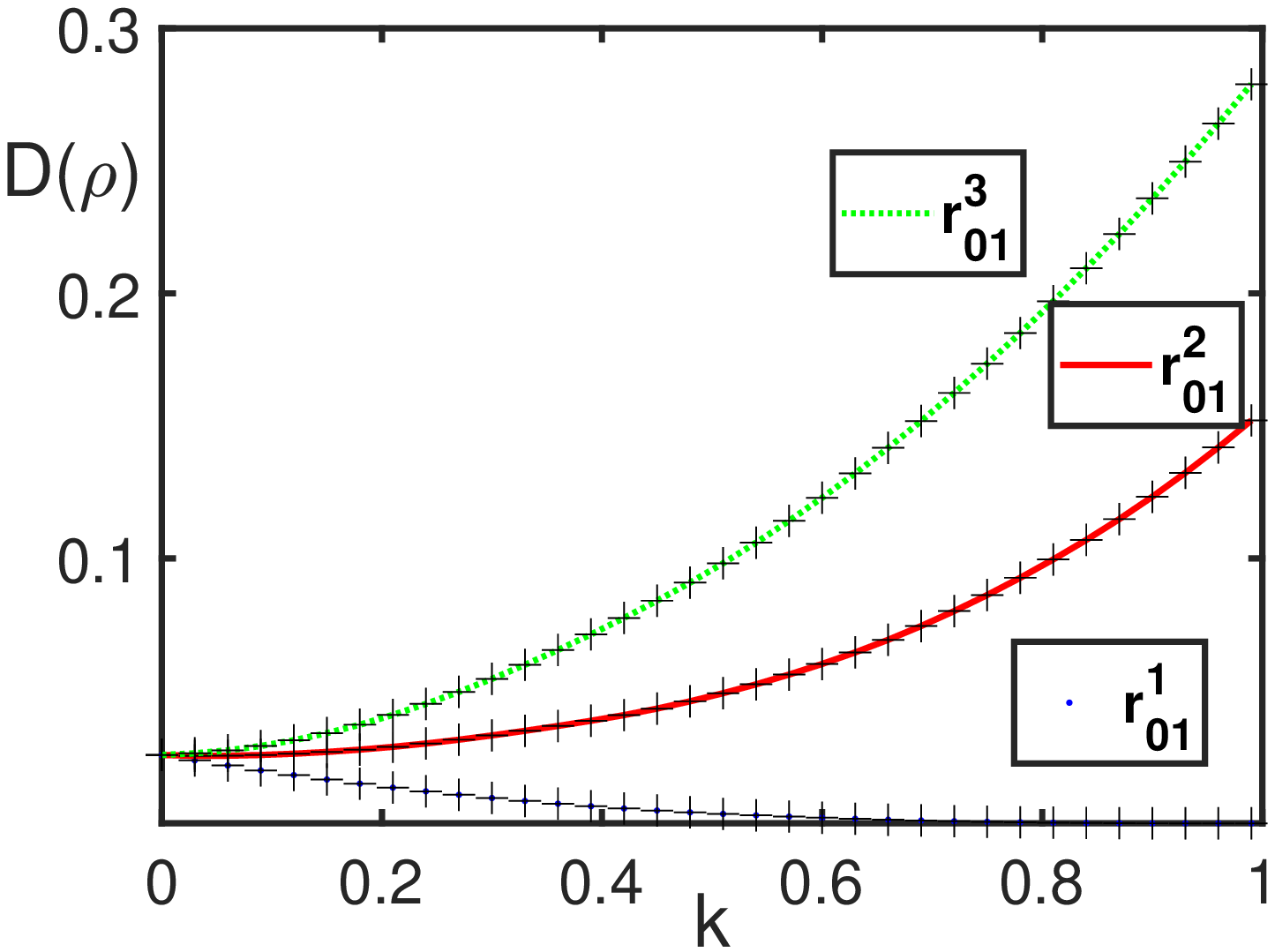}} %
\subfigure[]{\includegraphics[width=0.46\columnwidth,height=1.3in]{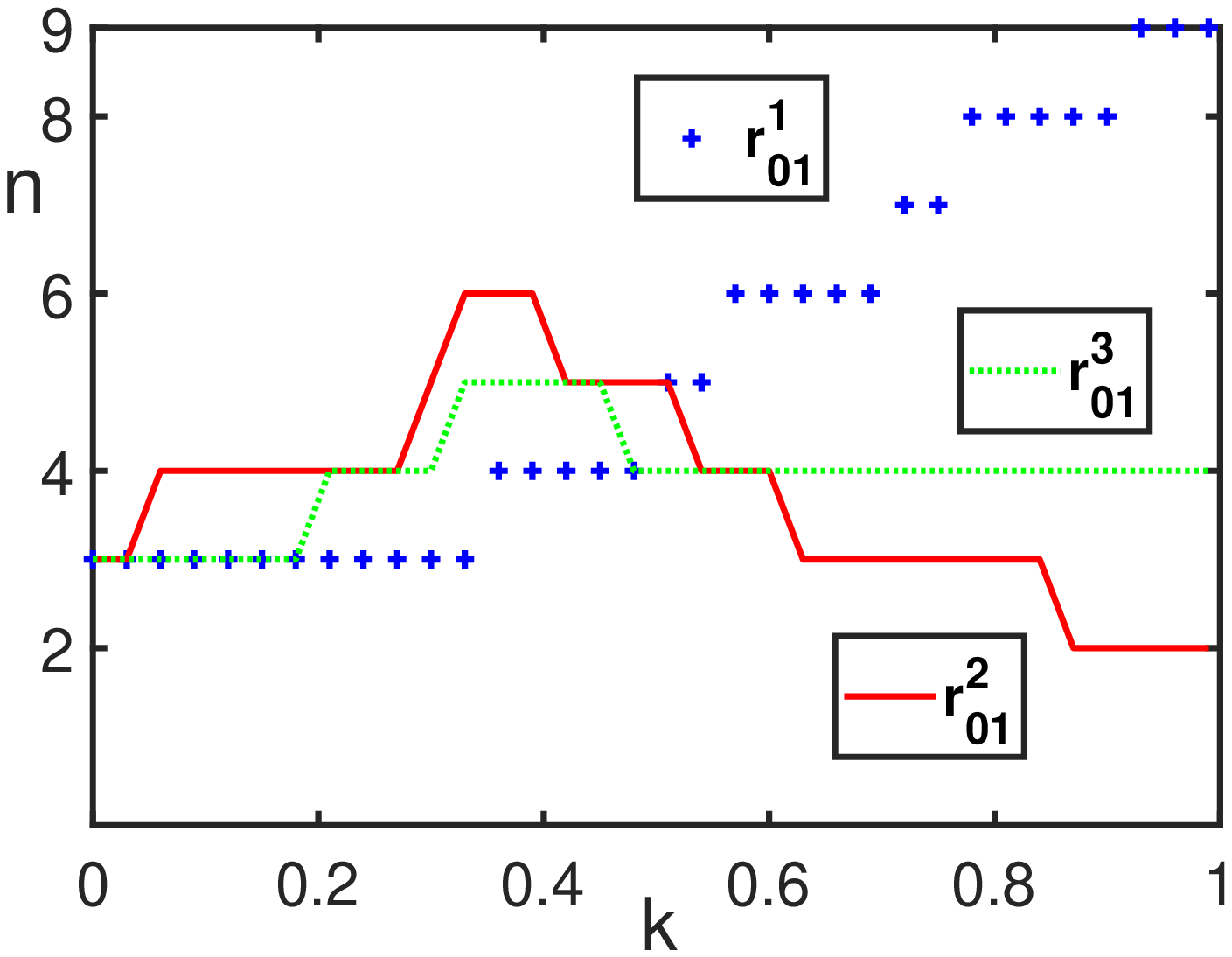}}%
\caption{(color online) The optimal distance $D(\protect\rho )$ versus
various parameters $k$ in (a) for $d=3$ and $N=15$. The solid line corresponds to the strictly
closed expressions, while the numerical solutions are marked with "+". The minimal number $n$ of quantum states needed for the optimal approximation of each objective state is shown in (b).}
\end{figure}

(iv)$d=4$ and $N=20$. Due to the large dimension of the considered states, the concrete expressions of  $\mathbf{r}_{01}$ and $\mathbf{r}_{02}$ are given in Appendix A. Similarly, we consider three special quantum states $\lbrace\mathbf{r}_{01}^{1},\mathbf{r}_{01}^{2},\mathbf{r}_{01}^{3}\rbrace$, and quantum state $\mathbf{r}_{01}^1$ is the maximally mixed state. The optimal distance denoted
by $D(\rho)$ versus $k\in \lbrack 0,1]$ is plotted in Fig. 4(a), which
shows the perfect consistency between the numerical and the closed
results, and further supports our theorem. As can be seen from Fig. 4 (b), the minimal number of quantum states in set $S$ used to approximate the target quantum state is less than or equal to 14. By comparing the figures, we can find that with the increase of the maximally mixed state ratio, the optimal approximate distance tends to 0.

\begin{figure}[tbp]
\centering
\subfigure[]{\includegraphics[width=0.50\columnwidth,height=1.3in]{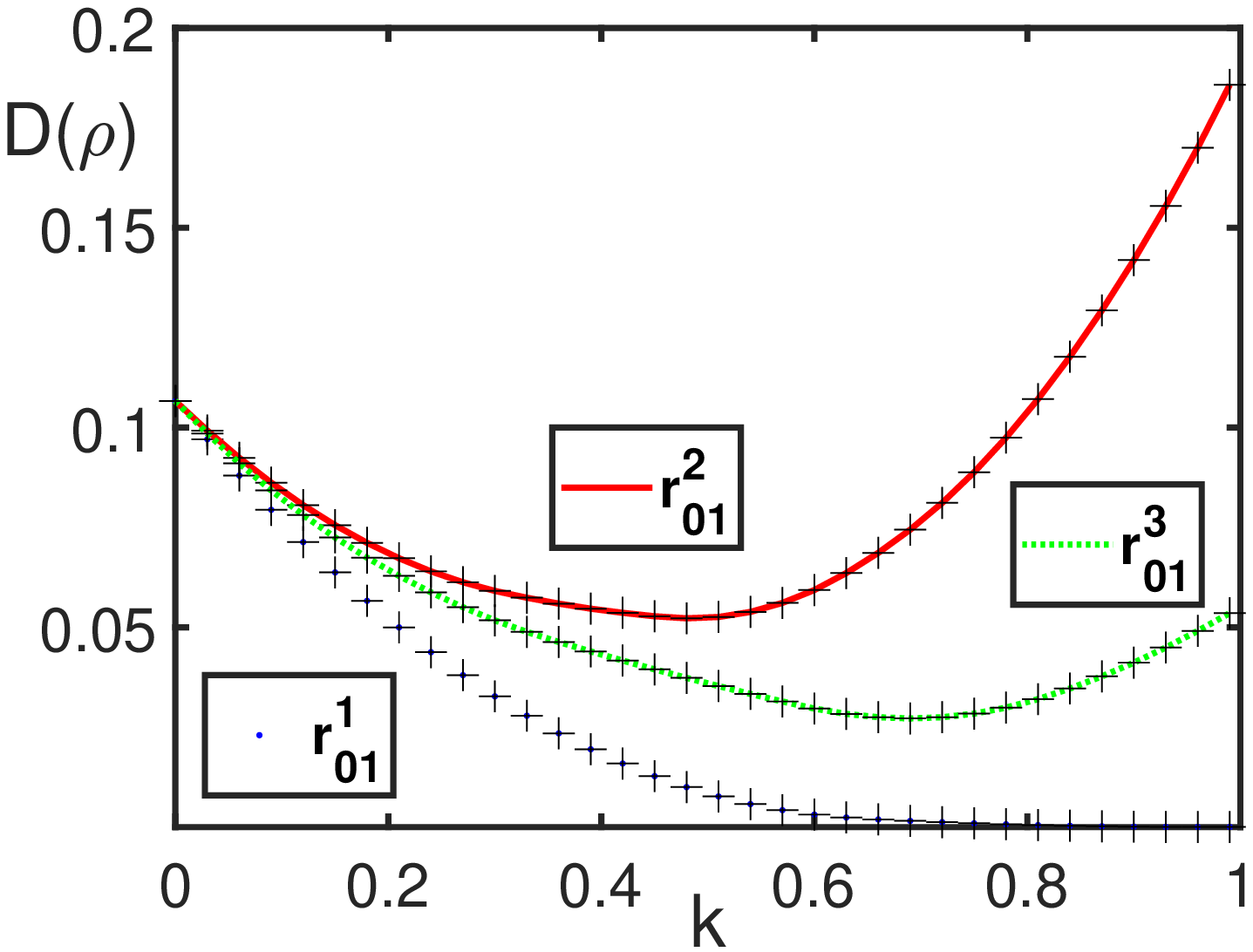}} %
\subfigure[]{\includegraphics[width=0.46\columnwidth,height=1.3in]{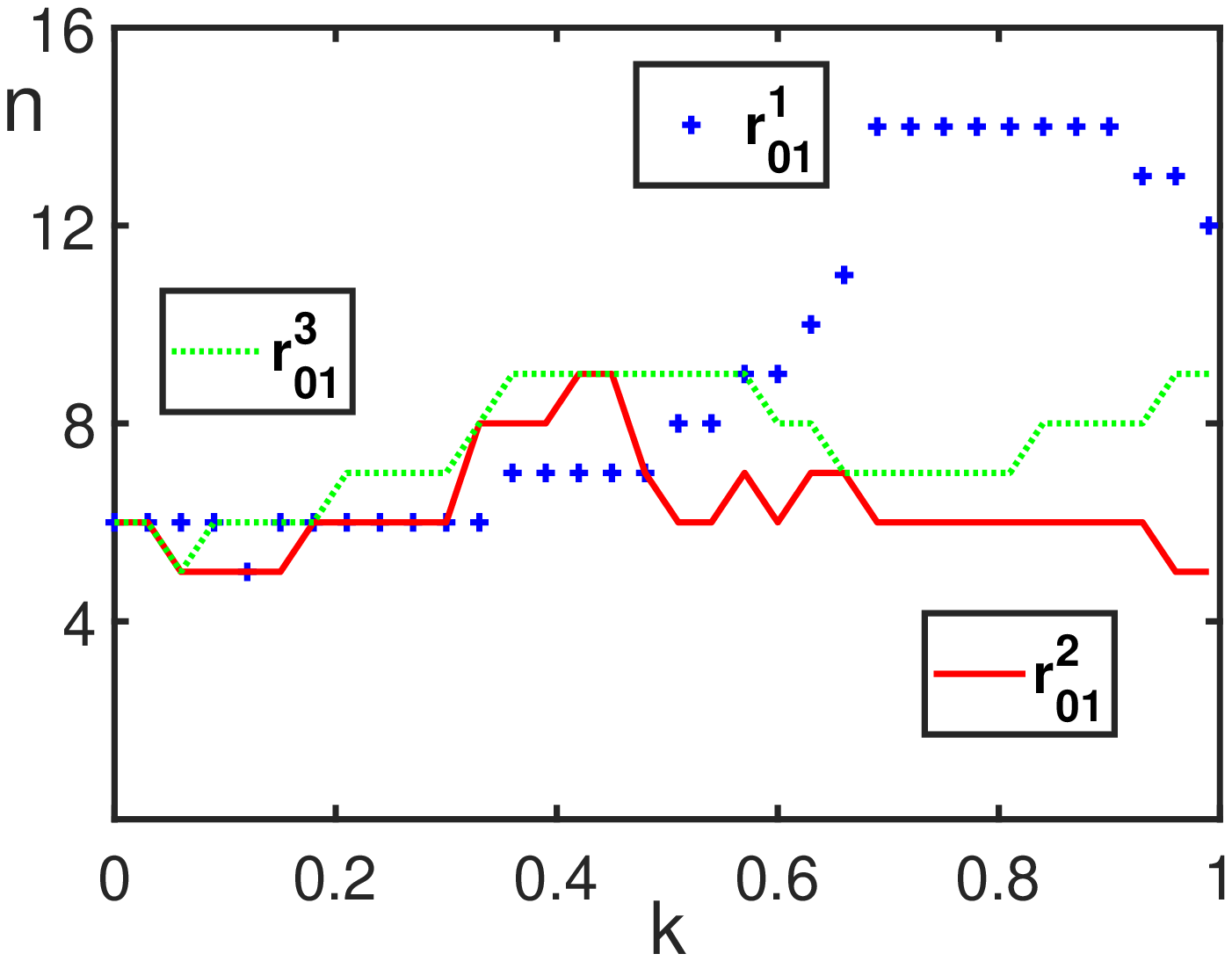}} %
\caption{(color online) The optimal distance $D(\protect\rho )$ versus
various parameters $k$ in (a) for $d=4$ and $N=20$. The solid line corresponds to the strictly
closed expressions in Eq. (\ref{theo3}), while the numerical solutions
are marked with "+". The minimal number $n$ of quantum states in set $S$ to optimally construct each objective state is shown in (b).}
\end{figure}

\section{Discussion and Conclusion}

Before the end, we'd like to mention that we have studied the best  approximation of an objective state by a limited state set based on $l_{2}$ norm, which provided a global convex optimization. The approach may be applied to other distance measures. If the given set $S$ is defined by $d$ mutually orthogonal quantum states, the optimal distance provides an alternative measure of quantum coherence of the objective quantum states based on $l_{2}$ norm, which is equal to the trace norm in the 2-dimensional case \cite{c1}. If $d$ linearly independent quantum states are given for the set $S$, our results provide a measure of the superposition of the objective states \cite{s1}. In addition, in the paper we only consider the approximation of a single party system, it is worth considering the local or nonlocal approximation of states in a composite system.

In summary, we have given a closed solution to the approximation of a $d$-dimensional objective state by using a given state set. We have not only presented the optimal distance between the objective state and the prepared state, but also given the minimal number of states in the given set $S$ to achieve the best approximation. Numerical test  in several examples validate our closed solutions via the perfect consistency. In addition, it is found that for the $d$-dimensional objectivce quantum state, the optimal distance can be achieved by the convex combination of no more than $d^2$ quantum states.
Finally, we emphasize that our closed solution indicates the least number of quantum states to construct the target quantum state approximately, which is beneficial to the practical operation in the experiment.

\appendix

\section{ The explicit forms of the set $S$ for examples}

\textit{In example} (\textit{i}), the quantum state set $S$ includes 3 quantum states, which can be expressed as
\begin{eqnarray}
\mathbf{r}_{1}&&=(\begin{array}{cccc}
1/\sqrt{2} ,& -0.0453  & -0.0429 &  -0.0774  %
\end{array}%
\mathbf{)^{\intercal }}, \notag\\
\mathbf{r}_{2}&&=(\begin{array}{cccc}
1/\sqrt{2} ,& -0.3348 &  -0.2708  & -0.2571   %
\end{array}%
\mathbf{)^{\intercal }}, \notag\\
\mathbf{r}_{3}&&=(\begin{array}{cccc}
1/\sqrt{2} ,& 0.0287  &  0.2456 &  -0.0534   %
\end{array}%
\mathbf{)^{\intercal }}.
\end{eqnarray}

\textit{In example} (\textit{ii}), the quantum state set $S$ includes 6 quantum states, which are composed of eigenstates of three Pauli matrix $\lbrace\sigma_{x},\sigma_{y},\sigma_{z}\rbrace$, expressed as
\begin{eqnarray}
\mathbf{r}_{1}&&=(\begin{array}{cccc}
1 ,& 1 ,& 0 ,& 0  %
\end{array}%
\mathbf{)^{\intercal }}/\sqrt{2}, \notag\\
\mathbf{r}_{2}&&=(\begin{array}{cccc}
1 ,& -1 ,& 0 ,& 0  %
\end{array}%
\mathbf{)^{\intercal }}/\sqrt{2}, \notag\\
\mathbf{r}_{3}&&=(\begin{array}{cccc}
1 ,& 0 ,& 1 ,& 0  %
\end{array}%
\mathbf{)^{\intercal }}/\sqrt{2}, \notag\\
\mathbf{r}_{4}&&=(\begin{array}{cccc}
1 ,& 0 ,& -1 ,& 0  %
\end{array}%
\mathbf{)^{\intercal }}/\sqrt{2}, \notag\\
\mathbf{r}_{5}&&=(\begin{array}{cccc}
1 ,& 0 ,& 0 ,& 1  %
\end{array}%
\mathbf{)^{\intercal }}/\sqrt{2}, \notag\\
\mathbf{r}_{6}&&=(\begin{array}{cccc}
1 ,& 0 ,& 0 ,& -1  %
\end{array}%
\mathbf{)^{\intercal }}/\sqrt{2}.
\end{eqnarray}

\textit{In example} (\textit{iii}), the quantum state set $S$ is given in operator Hilbert space as
\begin{widetext}
\begin{eqnarray}
   \mathbf{r}_{1}&&=(\begin{array}{ccccccccc}
  0.5774 ,&  -0.0089  ,&  0.0192  ,&  0.0446 ,&  -0.0585  ,& -0.0403  ,& -0.0061  ,&  0.0094   ,& 0.0210 %
\end{array}%
\mathbf{)^{\intercal }}, \notag \\
   \mathbf{r}_{2}&&=(\begin{array}{ccccccccc}
   0.5774 ,&  -0.0679 ,&  -0.0568 ,&   0.0278 ,&  -0.0335  ,&  0.0708  ,& -0.1094   ,& 0.1036  ,& -0.1595%
\end{array}%
\mathbf{)^{\intercal }}, \notag \\
   \mathbf{r}_{3}&&=(\begin{array}{ccccccccc}
 0.5774 ,&   0.3314  ,& -0.2469  ,& -0.0453 ,&  -0.0119 ,&  -0.2463  ,&  0.1383  ,&  0.0430  ,& -0.1808%
\end{array}%
\mathbf{)^{\intercal }}, \notag \\
   \mathbf{r}_{4}&&=(\begin{array}{ccccccccc}
  0.5774  ,&  0.0138  ,&  0.2443  ,& -0.2903  ,&  0.2369 ,&  -0.1108 ,&  -0.0694  ,&  0.2347  ,& -0.1709
\end{array}%
\mathbf{)^{\intercal }}, \notag \\
   \mathbf{r}_{5}&&=(\begin{array}{ccccccccc}
 0.5774 ,&  -0.1427  ,& -0.0667  ,& -0.4253  ,& -0.3187 ,&  -0.1495 ,&  -0.3435 ,&   0.3087   ,& 0.2761%
\end{array}%
\mathbf{)^{\intercal }},
\end{eqnarray}
%\end{widetext}
%\begin{widetext}
\begin{eqnarray}
   \mathbf{r}_{6}&&=(\begin{array}{ccccccccc}
 0.5774 ,&  -0.1204 ,&  -0.1235  ,&  0.0178 ,&  -0.1018 ,&  -0.0949 ,&   0.1963 ,&  -0.0295  ,& -0.0742%
\end{array}%
\mathbf{)^{\intercal }}, \notag \\
   \mathbf{r}_{7}&&=(\begin{array}{ccccccccc}
  0.5774 ,&  -0.2536  ,&  0.1011 ,&  -0.0765  ,&  0.2108  ,&  0.2307  ,&  0.3318 ,&  -0.1931   ,& 0.2599%
\end{array}%
\mathbf{)^{\intercal }}, \notag \\
   \mathbf{r}_{8}&&=(\begin{array}{ccccccccc}
 0.5774 ,&  -0.1104 ,&   0.1763 ,&  -0.0249 ,&   0.2200  ,&  0.3024 ,&  -0.0175  ,&  0.1947   ,& 0.2628%
\end{array}%
\mathbf{)^{\intercal }}, \notag \\
   \mathbf{r}_{9}&&=(\begin{array}{ccccccccc}
0.5774 ,&  -0.0702  ,&  0.1844 ,&  -0.1471  ,& -0.3248  ,&  0.2558  ,& -0.0720  ,& -0.3141  ,& -0.1375%
\end{array}%
\mathbf{)^{\intercal }}, \notag \\
   \mathbf{r}_{10}&&=(\begin{array}{ccccccccc}
  0.5774  ,&  0.0639  ,&  0.0988  ,&  0.1477 ,&  -0.1330  ,& -0.1263  ,&  0.1457  ,&  0.0014   ,& 0.1164%
\end{array}%
\mathbf{)^{\intercal }},
\end{eqnarray}
%\end{widetext}
and
%\begin{widetext}
\begin{eqnarray}
   \mathbf{r}_{11}&&=(\begin{array}{ccccccccc}
  0.5774  ,&  0.0480 ,&  -0.0487 ,&  -0.0103 ,&   0.0349 ,&   0.0063 ,&  -0.0174 ,&  -0.0342   ,& 0.0324%
\end{array}%
\mathbf{)^{\intercal }}, \notag \\
   \mathbf{r}_{12}&&=(\begin{array}{ccccccccc}
 0.5774  ,&  0.0888 ,&  -0.0107  ,&  0.0179  ,& -0.0196 ,&  -0.0241 ,&   0.0779 ,&   0.0117   ,& 0.1170%
\end{array}%
\mathbf{)^{\intercal }}, \notag \\
   \mathbf{r}_{13}&&=(\begin{array}{ccccccccc}
 0.5774 ,&  -0.0442 ,&  -0.0588  ,& -0.0470 ,&  -0.0180 ,&  -0.0025 ,&  -0.0006 ,&  -0.0301  ,& -0.0546%
\end{array}%
\mathbf{)^{\intercal }}, \notag \\
   \mathbf{r}_{14}&&=(\begin{array}{ccccccccc}
  0.5774  ,&  0.0104 ,&  -0.0121 ,&  -0.0060 ,&  -0.0082 ,&  -0.0087 ,&   0.0133 ,&   0.0029  ,& -0.0065%
\end{array}%
\mathbf{)^{\intercal }}, \notag \\
   \mathbf{r}_{15}&&=(\begin{array}{ccccccccc}
  0.5774  ,&  0.2361  ,& -0.2266 ,&   0.2744  ,&  0.1280  ,& -0.2198 ,&  0.2896 ,&  -0.1512   ,& 0.1012%
\end{array}%
\mathbf{)^{\intercal }}.
\end{eqnarray}
\end{widetext}

\textit{In example} (\textit{iv}), three special quantum states are
\begin{widetext}
\begin{eqnarray}
\mathbf{r}_{01}^{1}=&& \left( \begin{array}{cccccccccccccccc}
 0.5000  ,&  0  ,& 0  ,& 0 ,&  0 ,&  0  ,& 0 ,&   0,&0 ,&  0 ,&  0 ,&  0  ,&  0 ,& 0  ,&  0 ,&  0
\end{array}%
\right)^{\mathbf{\intercal}},\notag \\
\mathbf{r}_{01}^{2}=&& \left( \begin{array}{cccccccccccccccc}
1  ,&  \frac{1}{\sqrt{5}}   ,& \frac{1}{\sqrt{5}}   ,& \frac{1}{\sqrt{5}}  ,& \frac{1}{\sqrt{5}}  ,&  \frac{1}{\sqrt{5}}   ,& \frac{1}{\sqrt{5}}  ,&   \frac{1}{\sqrt{5}} ,&\frac{1}{\sqrt{5}}  ,&   \frac{1}{\sqrt{5}}  ,&  \frac{1}{\sqrt{5}}  ,&  \frac{1}{\sqrt{5}}   ,&  \frac{1}{\sqrt{5}}  ,&  \frac{1}{\sqrt{5}}   ,& \frac{1}{\sqrt{5}}  ,&  \frac{1}{\sqrt{5}} %
\end{array}%
\right)^{\mathbf{\intercal}}/2,\notag \\
\mathbf{r}_{01}^{3}=&& \left( \begin{array}{cccccccccccccccc}
1  ,&  \frac{1}{\sqrt{5}}   ,& \frac{1}{\sqrt{5}}   ,& \frac{1}{\sqrt{5}}  ,& \frac{1}{\sqrt{5}}  ,&  \frac{1}{\sqrt{5}}   ,& \frac{1}{\sqrt{5}}  ,&   \frac{1}{\sqrt{5}} ,&0  ,&   0  ,&  0  ,&  0   ,&  0  ,&  0  ,& 0  ,&  0%
\end{array}%
\right)^{\mathbf{\intercal}}/2,
\end{eqnarray}
%\end{widetext}
and the random generated quantum state $\mathbf{r}_{02}$ is
%\begin{widetext}
\begin{eqnarray}
\mathbf{r}_{02}=&& \left( \begin{array}{cccccccc}
 0.5000  ,&  0.3480  ,& -0.0903  ,& -0.2264 ,&  -0.0123 ,&  -0.3373  ,& -0.1569 ,&   0.2523,
\end{array} \right.  \notag \\
&& \left. \begin{array}{cccccccc}
0.0478 ,&   0.3409 ,&  -0.1137 ,&  -0.0728  ,&  0.0766 ,&  -0.3191  ,&  0.0155 ,&  -0.2063
\end{array}%
\right)^{\mathbf{\intercal}}.
\end{eqnarray}
\end{widetext}

In addition, the quantum state set $S$ includes 20 randomly generated quantum states which are
\begin{widetext}
\begin{eqnarray}
\mathbf{r}_{1}=&& \left( \begin{array}{cccccccc}
0.5000  ,&  0.3401 ,&   0.2281  ,&  0.1506 ,&  -0.1592  ,&  0.0518 ,&   0.1435  ,& -0.0227,
\end{array} \right.  \notag \\
&& \left. \begin{array}{cccccccc}
 0.0123 ,&  -0.0654 ,&  -0.1799  ,& -0.2337 ,&  -0.3263 ,&  -0.3152  ,& -0.3466  ,& -0.3052  %
\end{array}%
\right)^{\mathbf{\intercal}},\notag \\
\mathbf{r}_{2}=&& \left( \begin{array}{cccccccc}
  0.5000 ,&   0.2630  ,&  0.2397  ,& -0.0116   ,& 0.3371  ,& -0.0435 ,&  -0.2855  ,&  0.2873 ,
\end{array} \right.  \notag \\
&& \left. \begin{array}{cccccccc}
 0.3155  ,&  0.0002  ,& -0.0044   ,& 0.1938 ,&  -0.3148 ,&   0.2518 ,&  -0.0674  ,& -0.1989 %
\end{array}%
\right)^{\mathbf{\intercal}},\notag \\
\mathbf{r}_{3}=&& \left( \begin{array}{cccccccc}
  0.5000 ,&  -0.2184  ,& -0.0078 ,&  0.1946 ,&   0.1876  ,&  0.3716 ,&   0.2003  ,& -0.1040 ,
\end{array} \right.  \notag \\
&& \left. \begin{array}{cccccccc}
 -0.1933 ,&  -0.0483  ,&  0.0962  ,&  0.2114 ,&  -0.2066  ,&  0.4027 ,&  -0.3182 ,&  -0.2010 %
\end{array}%
\right)^{\mathbf{\intercal}},\notag \\
\mathbf{r}_{4}=&& \left( \begin{array}{cccccccc}
   0.5000 ,&  -0.3085 ,&   0.1438 ,&  -0.3556 ,&  -0.0276 ,&  -0.3483 ,&  -0.3359  ,&  0.1912  ,
\end{array} \right.  \notag \\
&& \left. \begin{array}{cccccccc}
 0.0149  ,& -0.0655 ,&   0.1375  ,&  0.2116  ,&  0.0990 ,&  -0.3137 ,&  -0.0414  ,&  0.2411 %
\end{array}%
\right)^{\mathbf{\intercal}},\notag \\
\mathbf{r}_{5}=&& \left( \begin{array}{cccccccc}
   0.5000 ,&  -0.2741 ,&  -0.1356 ,&   0.0618 ,&   0.2636 ,&  -0.1240  ,& -0.2177  ,&  0.1307,
\end{array} \right.  \notag \\
&& \left. \begin{array}{cccccccc}
 0.0534  ,&  0.3311  ,&  0.3185  ,&  0.3498  ,&  0.3522 ,&   0.1482 ,&  -0.0426  ,& -0.1383 %
\end{array}%
\right)^{\mathbf{\intercal}},
\end{eqnarray}
%\begin{widetext}
\begin{eqnarray}
\mathbf{r}_{6}=&& \left( \begin{array}{cccccccc}
 0.5000  ,&  0.1112  ,&  0.3263  ,&  0.3190 ,&  -0.1469 ,&  -0.3317  ,&  0.2964  ,& -0.2435,
\end{array} \right.  \notag \\
&& \left. \begin{array}{cccccccc}
 -0.0953 ,&  -0.0306  ,&  0.1241  ,&  0.3175 ,&  -0.2460 ,&   0.1503  ,&  0.1725  ,&  0.1075  %
\end{array}%
\right)^{\mathbf{\intercal}},\notag \\
\mathbf{r}_{7}=&& \left( \begin{array}{cccccccc}
 0.5000 ,&  -0.3397  ,&  0.1331  ,& -0.3639 ,&  -0.0465 ,&  -0.3013 ,&  -0.0081 ,&  -0.1089 ,
\end{array} \right.  \notag \\
&& \left. \begin{array}{cccccccc}
0.0448 ,&   0.2988   ,& 0.2088 ,&   0.1497 ,&   0.3078  ,&  0.3506  ,&  0.0583  ,&  0.0359 %
\end{array}%
\right)^{\mathbf{\intercal}},\notag \\
\mathbf{r}_{8}=&& \left( \begin{array}{cccccccc}
  0.5000  ,& -0.0234  ,& -0.1351 ,&  -0.0322  ,&  0.2409  ,& -0.3197  ,& -0.1314  ,& -0.2997 ,
\end{array} \right.  \notag \\
&& \left. \begin{array}{cccccccc}
 -0.3042  ,& -0.3151 ,&   -0.2814  ,& -0.0517 ,&  -0.0072 ,&   0.2808 ,&  -0.0275  ,&  0.3309 %
\end{array}%
\right)^{\mathbf{\intercal}},\notag \\
\mathbf{r}_{9}=&& \left( \begin{array}{cccccccc}
  0.5000  ,&  0.0011 ,&   0.0720  ,&  0.0828  ,&  0.3043 ,&   0.3365 ,&  -0.2482  ,& -0.2899  ,
\end{array} \right.  \notag \\
&& \left. \begin{array}{cccccccc}
-0.1409  ,&  0.1969  ,&  0.2455  ,& -0.1839  ,&  0.2696 ,&  -0.2972  ,& -0.2457  ,&  0.1112 %
\end{array}%
\right)^{\mathbf{\intercal}},\notag \\
\mathbf{r}_{10}=&& \left( \begin{array}{cccccccc}
  0.5000 ,&  -0.0969 ,&  -0.2549  ,& -0.1631 ,&   0.3248 ,&  -0.1620 ,&  -0.2249  ,&  0.0524,
\end{array} \right.  \notag \\
&& \left. \begin{array}{cccccccc}
 -0.3021  ,& -0.2719  ,&  0.2176 ,&  -0.3062 ,&  -0.1522 ,&  -0.0556 ,&  -0.1959 ,&  -0.3050 %
\end{array}%
\right)^{\mathbf{\intercal}},
\end{eqnarray}
%\end{widetext}
%\begin{widetext}
\begin{eqnarray}
\mathbf{r}_{11}=&& \left( \begin{array}{cccccccc}
  0.5000  ,& -0.1085 ,&  -0.1859 ,&  -0.2159 ,&  -0.3368 ,&  -0.1676 ,&  -0.2280  ,&  0.2087,
\end{array} \right.  \notag \\
&& \left. \begin{array}{cccccccc}
 -0.0767  ,&  0.2550 ,&  -0.2157  ,& -0.2451 ,&  -0.0837 ,&  -0.3492 ,&  -0.0487  ,& -0.3334  %
\end{array}%
\right)^{\mathbf{\intercal}},\notag \\
\mathbf{r}_{12}=&& \left( \begin{array}{cccccccc}
 0.5000 ,&  -0.1116  ,&  0.1404 ,&  -0.3683 ,&  -0.2396 ,&   0.2121 ,&  0.0992 ,&  -0.3009 ,
\end{array} \right.  \notag \\
&& \left. \begin{array}{cccccccc}
0.0709 ,&  -0.3062  ,& -0.2525 ,&   0.2682  ,&  0.0603 ,&  -0.3460 ,&  -0.0274  ,& -0.1445 %
\end{array}%
\right)^{\mathbf{\intercal}},\notag \\
\mathbf{r}_{13}=&& \left( \begin{array}{cccccccc}
 0.5000 ,&   0.2993 ,&  -0.2373 ,&  -0.1311 ,&  -0.0875 ,&  -0.1534  ,&  0.1559  ,&  0.1657 ,
\end{array} \right.  \notag \\
&& \left. \begin{array}{cccccccc}
-0.2268  ,&  0.3667 ,&  -0.1520 ,&  0.0617 ,&  -0.1243 ,&   0.3017 ,&  -0.2709 ,&  -0.3336 %
\end{array}%
\right)^{\mathbf{\intercal}},\notag \\
\mathbf{r}_{14}=&& \left( \begin{array}{cccccccc}
 0.5000  ,&  0.3539 ,&   0.0031 ,&  -0.3141  ,& -0.1757 ,&  -0.3098 ,&   0.3047 ,&  -0.0659  ,
\end{array} \right.  \notag \\
&& \left. \begin{array}{cccccccc}
0.1496 ,&  -0.0462 ,&  -0.0051 ,&  -0.2103 ,&   0.2230  ,& -0.1621  ,&  0.3111  ,&  0.2460 %
\end{array}%
\right)^{\mathbf{\intercal}},\notag \\
\mathbf{r}_{15}=&& \left( \begin{array}{cccccccc}
  0.5000  ,&  0.1284  ,&  0.3134  ,&  0.2846  ,&  0.1660 ,&  -0.0162  ,& -0.0902  ,&  0.2459 ,
\end{array} \right.  \notag \\
&& \left. \begin{array}{cccccccc}
 -0.3016 ,&   0.0968 ,&  -0.2279  ,&  0.2806 ,&  -0.1189  ,&  0.2264  ,& -0.3173 ,&  -0.2467 %
\end{array}%
\right)^{\mathbf{\intercal}},
\end{eqnarray}
%\end{widetext}
and
%\begin{widetext}
\begin{eqnarray}
\mathbf{r}_{16}=&& \left( \begin{array}{cccccccc}
  0.5000  ,& -0.0621 ,&   0.1726 ,&  -0.0117 ,&   0.1752  ,&  0.4750  ,&  0.1905  ,&  0.3538,
\end{array} \right.  \notag \\
&& \left. \begin{array}{cccccccc}
 -0.0945 ,&   0.0778 ,&  -0.0987 ,&  -0.3450 ,&  -0.0469 ,&  -0.0181  ,&  0.2563  ,&  0.2941  %
\end{array}%
\right)^{\mathbf{\intercal}},\notag \\
\mathbf{r}_{17}=&& \left( \begin{array}{cccccccc}
 0.5000 ,&  -0.3708 ,&  -0.1331 ,&  -0.2681  ,&  0.2228 ,&  -0.0237 ,&   0.0812 ,&  -0.3113,
\end{array} \right.  \notag \\
&& \left. \begin{array}{cccccccc}
 -0.2672 ,&  -0.2705 ,&   0.1829 ,&   0.0050 ,&  -0.1646 ,&  -0.1872  ,&  0.3579  ,&  0.0310 %
\end{array}%
\right)^{\mathbf{\intercal}},\notag \\
\mathbf{r}_{18}=&& \left( \begin{array}{cccccccc}
  0.5000  ,&  0.1800  ,&  0.1847 ,&  -0.2499 ,&  -0.1384  ,&  0.3771  ,&  0.3414  ,& -0.0918  ,
\end{array} \right.  \notag \\
&& \left. \begin{array}{cccccccc}
 -0.2475  ,&  0.0291  ,& -0.2798  ,& -0.1287 ,&  -0.2310  ,& -0.1423 ,&  -0.2583 ,&  -0.1931 %
\end{array}%
\right)^{\mathbf{\intercal}},\notag \\
\mathbf{r}_{19}=&& \left( \begin{array}{cccccccc}
 0.5000 ,&  -0.2897  ,&  0.0626 ,&  -0.2482  ,&  0.0671 ,&   0.2700  ,&  0.2767 ,&  -0.0325  ,
\end{array} \right.  \notag \\
&& \left. \begin{array}{cccccccc}
0.2770 ,&  -0.2419 ,&   0.0868 ,&  -0.2809 ,&   0.1914  ,&  0.2186 ,&  -0.2852 ,&  -0.2412 %
\end{array}%
\right)^{\mathbf{\intercal}},\notag \\
\mathbf{r}_{20}=&& \left( \begin{array}{cccccccc}
 0.5000  ,&  0.2981 ,&  -0.3500  ,&  0.0817  ,& -0.1222 ,&  -0.4227 ,&  -0.1678  ,&  0.1315,
\end{array} \right.  \notag \\
&& \left. \begin{array}{cccccccc}
 0.3039  ,&  0.0275  ,&  0.2742  ,& -0.0520 ,&   0.1699  ,&  0.2849  ,&  0.0298 ,&   0.1050 %
\end{array}%
\right)^{\mathbf{\intercal}}.
\end{eqnarray}
\end{widetext}

\acknowledgments This work was supported by the National Natural Science Foundation of China under Grant
No. 11775040, No. 12011530014, the Fundamental Research Fund for the Central Universities under Grant No. DUT20LAB203, and the Key Research and Development Project of Liaoning Province under Grant No.2020JH2/10500003.

\end{document}